\documentclass[11pt,onecolumn]{article}
\usepackage[utf8]{inputenc}
\usepackage{amsmath,amsfonts,amssymb,graphicx,authblk,placeins,cite}
\usepackage{comment}
\usepackage{color,soul}
\usepackage{hyphenat}
\usepackage[top=50pt,bottom=50pt,left=70pt,right=70pt]{geometry}
\usepackage{sectsty}
\usepackage[font=sf]{caption}

\begin{document}
\allsectionsfont{\sffamily}
\pagenumbering{gobble}

\title{\sffamily \Huge Towards a Cognitive Computational Neuroscience of Auditory Phantom Perceptions
 \vspace{1 cm}}


\author[1,2,3,4,*]{\sffamily Patrick Krauss}
\author[1,2,5,*]{\sffamily Achim Schilling}


\affil[1]{\small{Neuroscience Lab, University Hospital Erlangen, Germany}}

\affil[2]{\small{Cognitive Computational Neuroscience Group, Chair of English Philology and Linguistics, Friedrich-Alexander University Erlangen-N\"urnberg (FAU), Germany}}

\affil[3]{\small Linguistics Lab, University Erlangen-N\"urnberg, Germany}

\affil[4]{\small Cognitive Neuroscience Center, University of Groningen, The Netherlands}

\affil[5]{\small Laboratory of Sensory and Cognitive Neuroscience, Aix‐Marseille University, Marseille, France}

\affil[ ]{\small }

\affil[*]{\small both authors contributed equally}

\maketitle


\begin{abstract}{\sffamily \noindent
In order to gain a mechanistic understanding of how tinnitus emerges in the brain, we must build biologically plausible computational models that mimic both tinnitus development and perception, and test the tentative models with brain and behavioral experiments. With a special focus on tinnitus research, we review recent work at the intersection of artificial intelligence, psychology and neuroscience, indicating a new research agenda that follows the idea that experiments will yield theoretical insight only when employed to test brain-computational models. This view challenges the popular belief, that tinnitus research is primarily data limited, and that producing large, multi-modal, and complex datasets, analyzed with advanced data analysis algorithms, will finally lead to fundamental insights into how tinnitus emerges. However, there is converging evidence that although modern technologies allow assessing neural activity in unprecedentedly rich ways in both, animals and humans, empirical testing one verbally defined hypothesis about tinnitus after another, will never lead to a mechanistic understanding. Instead, hypothesis testing needs to be complemented with the construction of computational models that generate verifiable predictions. We argue, that even though, contemporary artificial intelligence and machine learning approaches largely lack biological plausibility, the models to be constructed will have to draw on concepts from these fields, since they have already proven to do well in modeling brain function. Nevertheless, biological fidelity will have to be increased successively, leading to ever better and fine-grained models, allowing at the end for even testing possible treatment strategies in silico, before application in animal or patient studies.
}
\end{abstract}

\newpage
\section*{Introduction}

What is the ultimate goal of tinnitus research? -- To gain a mechanistic understanding of how tinnitus emerges in the brain, which means in detail, how the neural and mental processes underlying perception, cognition and behavior contribute to and are affected by the development of tinnitus, and finally, how the generation of tinnitus may be reversed or at least mitigated. While there is broad agreement in the community on the goal, there is far less agreement on the way to achieve it. There is still a popular belief among neuroscientific and psychological tinnitus researchers that we are primary data limited, and that producing large, multi-modal, and complex data sets, analyzed with advanced data analysis algorithms will finally lead to fundamental insights into how tinnitus emerges in the brain. This is known as the so called data-driven or bottom-up approach, respectively. However, this view is strongly challenged by a number of theoretical considerations, thought experiments, and empirical in silico studies, which are briefly summarized in the following sections.

\section*{Can a tinnitus researcher fix a radio?}

Already in 2002, Yuri Lazebnik compared biologists' endevour of trying to understand the building blocks and processes of living cells with problems that engineers typically deal with \cite{lazebnik2002can}. In his paper \emph{Can a biologist fix a radio? -- Or, what I learned while studying apoptosis}, Lazebnik argued that many fields of biomedical research at some point reach \emph{``a stage at which models, that seemed so complete, fall apart, predictions that were considered so obvious are found to be wrong, and attempts to develop wonder drugs largely fail. This stage is characterized by a sense of frustration at the complexity of the process''} \cite{lazebnik2002can}.

Subsequently, Lazebnik discussed a number of intriguing analogies between engineering and life sciences. In particular, he identified formal language as the most important difference between the two disciplines. Lazebnik argues that biologists and engineers use very different languages for describing phenomena. On the one hand, biologists draw box-and-arrow diagrams, which are - even if a certain diagram makes overall sense - usually useless for a quantitative analysis, and hence limits its predictive or investigative value to a very narrow range. Also for verbal communications, according to Lazebnik, biologists use a language that is neither better, nor unlike to the language used by stock market analysts. He argues that both are \emph{``vague and avoid clear, quantifiable predictions''} \cite{lazebnik2002can}. A freely adapted example drawn from Lazebnik's paper would be a statement like \emph{``an imbalance of excitatory and inhibitory neural activity after hearing-loss appears to cause an overall neural hyperactivity, which in turn seems to be correlated with the perception of tinnitus.''}

Lazebnik futher argues that these tools frequently used in life sciences for description and communication are in a glaring contrast with the formal language used in engineering or physics. Here, the language is standardized, i.e. the elements, their connections and interactions are described according to invariable rules. As a consequence, any engineer trained in electronics for instance, is able to unambiguously understand a diagram describing a radio or any other electronic device. Thus, engineers can discuss a radio (or physicists electrodynamics, respectively) using terms that are understood unambiguously in the community. Furthermore, this commonality of their language enables engineers to identify familiar patterns or modules even in a diagram of a completely unfamiliar device. Finally, due to the quantitative nature of the language used in engineering, it is perfectly suited for quantitative analyses and computational modeling. For instance, a description of a certain radio includes all key parameters of each component like the capacity of a capacitor, but not inadequate parameters like its color, shape, or size, being meaningless for understanding the design principles of the radio.

Lazebnik concludes that \emph{``the absence of such language is the flaw of biological research that causes David’s paradox''}, i.e. the paradox phenomenon frequently observed in biology and neuroscience that \emph{``the more facts we learn the less we understand the process we study''} \cite{lazebnik2002can}. Some experimentalists nevertheless may ask why formal language should be taken care of if one can build a scientific career on purely experimental research. Lazebnik argues that there are at least two reasons. Firstly, formal approaches make research more productive, empirical results more meaningful, and might indeed lead to completely new treatment strategies. Secondly, more immediate, formal approaches may soon become a such important part of life sciences and biomedical research that this transition may be as disruptive as that from slides to PowerPoint presentations. Lazebnik ends his considerations with the advice to experimental researchers to better be prepared.

\section*{The tale of the tinnitus researchers and the computer: \\ Why mechanistic theory matters}

In 2014, Joshua Brown built on Lazebnik’s ideas and published the opinion article \emph{The tale of the neuroscientists and the computer: why mechanistic theory matters} \cite{brown2014tale}. In this thought experiment, a group of neuroscientists finds an alien computer and tries to figure out its function.

First, the M/EEG researcher tried her luck. She found, that every time \emph{``when the hard disk was assessed, the disk controller showed higher voltages on average, and especially more power in the higher frequency bands''} \cite{brown2014tale}. Subsequently, the cognitive neuroscientist, i.e. the fMRI researcher, came up and argued that M/EEG has not enough spatial resolution to see what is going on inside the computer. He carried out a large number of experiments, the results of which can be summarized with the realization that during certain tasks, certain regions seem to be more activated. Eventually, he had the crucial insight that none of these components could be understood properly in isolation. Instead, researchers had to understand the networks. Again, a bunch of experiments was carried out, showing that there is a vast variety of different task-specific networks in the computer. They even found a very special network not associated to any certain task, but instead indicating that the computer is idle.

Finally, the electrophysiologist spoke up and argued that his colleagues may have found the larger patterns of activity, but it is still unclear what the individual circuits are doing. He starts to implant microelectrode arrays into the computer and probes individual circuit points within the computer by measuring the time course of the voltage. With careful observation, the electrophysiologist identifies units responding stochastically when certain inputs are presented, and that nearby units seem to process similar inputs. Furthermore, each unit seems to have characteristic tuning properties.

Brown’s tale ends with the conclusion that even though, they performed a multitude of different empirical investigations, yielding a broad range of interesting results, it is still highly questionable whether \emph{``the neuroscientists really understood how the computer works''} \cite{brown2014tale}.

\section*{Could a tinnitus researcher understand a microprocessor?}

In their seminal study, Eric Jonas and Konrad Kording put the aforementioned thoughts even one step further, and ask the provoking question \emph{Could a neuroscientist understand a microprocessor?} \cite{jonas2017could}. The authors address this question by emulating a classical microprocessor, the \emph{MOS 6502}, which was implemented as the central processing unit (CPU) in the Apple I, the Commodore 64, and the Atari Video Game System, in the 1970s and 1980s. In contrast to contemporary CPUs, like Intel’s \emph{i9-9900K}, that consist of more than 3 billion transistors, the \emph{MOS 6502} only consisted of 3,510 transistors, served as a ``model organism'' in the mentioned study, and performed three different ``behaviors'', i.e. three classical video games (Donkey Kong, Space Invaders and Pitfall).

The idea behind this approach is that the microprocessor, as an artificial information processing system, has three decisive advantages over natural nervous systems. Firstly, it is fully understood at all levels of description and complexity, from its gross architecture and the overall data flow, through logical gate primitives, to the dynamics of single transistors. Secondly, its internal state is fully accessible without any restrictions to temporal or spatial resolution. And thirdly, it provides the ability to perform arbitrary experiments on it. Using this framework, the authors applied a wide range of popular data analysis methods from neuroscience to investigate the structural and dynamical properties of the microprocessor. The methods used included but were not restricted to Granger causality for analyzing task specific functional connectivity, time-frequency analysis as a hallmark of M/EEG research, spike pattern statistics, dimensionality reduction, lesions, and tuning curve analysis.

However, the results were extremely sobering and a shock to the research community. Although, each of the applied methods yielded results strikingly similar to what is known from neuroscientific or psychological studies, none of them could actually elucidate how the microprocessor works, or more broadly speaking, was appropriate to gain a mechanistic understanding of the investigated system.

\section*{What does it mean to understand a system?}

If popular analysis methods fail in gaining a mechanistic understanding, which alternative approaches are appropriate then? -- Most obviously, hypotheses about the structure and function of the system under investigation could have helped. Instead of simply collecting data with the available measurement methods and then evaluating the data with standard or whatever sophisticated analysis methods in the hope of learning something about the functioning of the system under investigation, it would be much more effective to have a concrete hypothesis about structure or function of the system and then to test it. Ideally, this hypothesis would be derived from an underlying theory. As Kurt Lewin, father of modern experimental psychology, pointed out: \emph{``There is nothing so practical as a good theory''} \cite{lewin1951field}. If we had theorized that the microprocessor performs arithmetic calculations, we could have derived the hypothesis, for example, that there must be something like 1-bit adders, and could have searched for them specifically.

However, Allan Newell, one of the fathers of artificial intelligence, stated that \emph{``You can‘t play 20 questions with nature and win''} \cite{newell1973you}. This means that testing one verbally defined hypothesis about a system or phenomenon after another will never lead to a mechanistic understanding. So, this raises the fundamental question of what it actually means to really understand a system.  

Yuri Lazebnik argued that understanding of a system was achieved when one could fix a broken implementation: \emph{``Understanding of a particular region or part of a system would occur when one could describe so accurately the inputs, the transformation, and the outputs that one brain region could be replaced with an entirely synthetic component''} \cite{lazebnik2002can}. 

According to David Marr, one can seek to understand a system at differing, complementary levels of analysis \cite{marr1979computational}. He distinguished the computational, the algorithmic and the implementational level of analysis.
The \emph{computational level} is the most coarse-grained level of analysis. It addresses the question of what is the problem or the task the system is seeking to solve via computation, resulting in the observed phenomena, e.g. behavior and perception, in our context in particular, phantom perceptions like tinnitus. This level of analysis is broadly addressed by the fields of psychology and cognitive neuroscience.
In contrast, the \emph{implementational level} represents the most fine-grained description of a system. Here, the system's concrete, physical layout is analyzed. In computer science and engineering, this corresponds to the exact hardware architecture and the individual software realization with a particular programming language. In the brain, where there exists no clear distinction between software and hardware, this level of description corresponds to the structural design of ion channels, synapses, neurons, local circuits and larger systems, and the physiological processes these components are subject to. This level of analysis can be considered as the hallmark of physiology and neurobiology.
Finally, the \emph{algorithmic level} takes an intermediate position between the previously described levels. It is about which algorithms, physically realized as described at the implementational level, the system employs to manipulate its internal representations, in order to solve the tasks and problems identified at the computational level. In computer science, the algorithmic level would be described independently of a specific programming language by abstract pseudo code.

We argue, that analysis at the algorithmic level is most crucial to understand auditory phantom perceptions like tinnitus or Zwicker tone. Only by knowing the algorithms that underlie normal auditory perception, we will gain a detailed understanding of what exactly happens under certain unhealthy conditions such as hearing loss and which processes eventually cause the development of tinnitus, so that we can eventually counteract these processes. 

Which discipline addresses this level of analysis in tinnitus research? Computational neuroscience comes to mind immediately. However, in 'good old-fashioned' computational neuroscience, great efforts have been made to model the physiological processes at the level of single neurons, dendrites, axons, synapses or even ion channels, leading to increasingly complex computational models. These models, mostly based on systems of coupled differential equations, are able to mimic experimental data in great detail. Perhaps the most popular among these models is the famous Hodgkin-Huxley model \cite{hodgkin1952quantitative}, which reproduces the temporal course of the membrane potential of a single neuron with impressive accuracy. These types of models are of great importance to deepen our understanding of fundamental physiological processes. However, in our opinion, they must also be considered as belonging to the implementational level of analysis, since they merely describe the physical realization of the algorithms, rather than the algorithms themselves.

In the following section, we will discuss which emerging research direction is the most promising candidate for being associated to the algorithmic level of analysis in the context of tinnitus research.

\begin{figure}[ht!]
	\centering
	\includegraphics[width=1.0\linewidth]{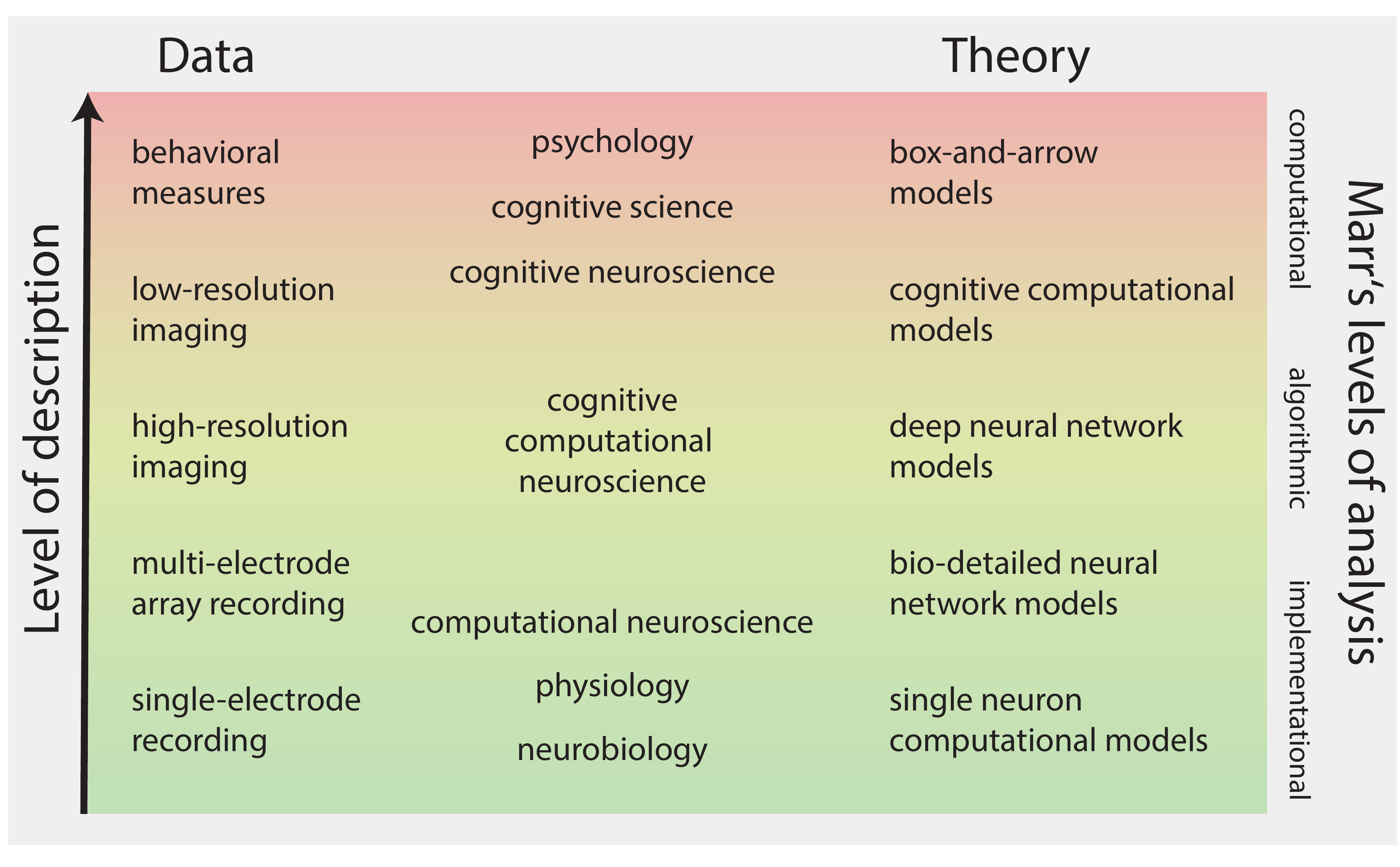}
	\caption{\textbf{Levels of description and analysis.}} 
	\label{descriptionLevels}
\end{figure}

\section*{Discussion: Towards a Cognitive Computational Neuroscience of Tinnitus}

As we argued above, hypothesis testing alone does not lead to a mechanistic understanding. Instead, it needs to be complemented by the construction of task-performing computational models since only synthesis in a computer simulation can reveal what the interaction of the proposed component mechanisms actually entails, i.e. which algorithms are realized, and whether they can account for the perceptual, cognitive or behavioral function in question. As Nobel laureate and theoretical physicist Richard Feynman pointed out: \emph{``What I cannot create, I do not understand.''}

Along these lines, one may consider to extend the traditional four goals of psychology, -- i.e. describing, explaining, predicting, and changing cognition and behavior, -- by adding a fifth one: building artificial cognition and behavior.

As pointed out in previous publications \cite{marblestone2016toward, van2017computational, van2017artificial, barak2017recurrent, kriegeskorte2018cognitive}, these computational models will have to be based on concepts of artificial intelligence and machine learning, in particular deep learning \cite{lecun2015deep, schmidhuber2015deep}, at least at the beginning. Artificial deep neural networks are designed to solve problems clearly defined at the computational level of analysis, in our case auditory perception tasks like, e.g. speech recognition. Moreover, these models are already precisely defined at an algorithmic level which is completely independent from any individual programming language or specific software library, i.e. the implementational level of analysis. Hence, these algorithms could, at least in principle, also be realized in the brain as biological neural networks. Once we have built such models and algorithms in computer simulations, we can subsequently compare their dynamics and internal representations with brain and behavioral data in order to reject or adjust putative candidate models, thereby successively increasing biological fidelity \cite{kriegeskorte2018cognitive}. Vice versa, the constructed models may also serve to generate new testable hypotheses about cognitive and neural processing in auditory neuroscience. 

For this research approach of combining artificial intelligence, cognitive science, and neuroscience, the term \emph{Cognitive Computational Neuroscience} has been coined \cite{kriegeskorte2018cognitive}.

Furthermore, besides the discussed advantages, this approach opens up the opportunity of in silico testing of new, putative treatment strategies for conditions like tinnitus prior to in vivo experiments. By that, Cognitive Computational Neuroscience may even serve to reduce the number of animal experiments.

However, we want to mention that Cognitive Computational Neuroscience of auditory perception is not only beneficial for neuroscience. As Hassabis pointed out, understanding biological brains could play a vital role in building intelligent machines, and that current advances in artificial intelligence have been inspired by the study of neural computation in humans and animals \cite{hassabis2017neuroscience}. Thus, Cognitive Computational Neuroscience of auditory perception, may contribute to the development of neuroscience-inspired artificial intelligence systems in the domain of natural language processing \cite{cambria2014jumping}. Finally, neuroscience may even serve to investigate \emph{machine behavior} \cite{rahwan2019machine}, i.e. shed light into the black box of deep learning \cite{voosen2017ai, hutson2018artificial}.

\section*{Acknowledgments}

This work was funded by the Deutsche Forschungsgemeinschaft (DFG, German Research Foundation): grant KR5148/2-1 to PK -- project number 436456810, and the Emerging Talents Initiative (ETI) of the University Erlangen-Nuremberg (grant 2019/2-Phil-01 to PK).

\section*{Author contributions}
PK and AS contributed equally to this work.

\section*{Competing interests}
The authors declare no competing financial interests.

\newpage
\FloatBarrier


\begin{thebibliography}{10}

\bibitem{lazebnik2002can}
Yuri Lazebnik.
\newblock Can a biologist fix a radio?—or, what i learned while studying
  apoptosis.
\newblock {\em Cancer cell}, 2(3):179--182, 2002.

\bibitem{brown2014tale}
Joshua~W Brown.
\newblock The tale of the neuroscientists and the computer: why mechanistic
  theory matters.
\newblock {\em Frontiers in neuroscience}, 8:349, 2014.

\bibitem{jonas2017could}
Eric Jonas and Konrad~Paul Kording.
\newblock Could a neuroscientist understand a microprocessor?
\newblock {\em PLoS computational biology}, 13(1):e1005268, 2017.

\bibitem{lewin1951field}
Kurt Lewin.
\newblock Field theory in social science: selected theoretical papers (edited
  by dorwin cartwright.).
\newblock 1951.

\bibitem{newell1973you}
Allen Newell.
\newblock You can't play 20 questions with nature and win: Projective comments
  on the papers of this symposium.
\newblock 1973.

\bibitem{marr1979computational}
David Marr and Tomaso Poggio.
\newblock A computational theory of human stereo vision.
\newblock {\em Proceedings of the Royal Society of London. Series B. Biological
  Sciences}, 204(1156):301--328, 1979.

\bibitem{hodgkin1952quantitative}
Alan~L Hodgkin and Andrew~F Huxley.
\newblock A quantitative description of membrane current and its application to
  conduction and excitation in nerve.
\newblock {\em The Journal of physiology}, 117(4):500, 1952.

\bibitem{marblestone2016toward}
Adam~H Marblestone, Greg Wayne, and Konrad~P Kording.
\newblock Toward an integration of deep learning and neuroscience.
\newblock {\em Frontiers in computational neuroscience}, 10:94, 2016.

\bibitem{van2017computational}
Marcel Van~Gerven.
\newblock Computational foundations of natural intelligence.
\newblock {\em Frontiers in computational neuroscience}, 11:112, 2017.

\bibitem{van2017artificial}
Marcel Van~Gerven and Sander Bohte.
\newblock Artificial neural networks as models of neural information
  processing.
\newblock {\em Frontiers in Computational Neuroscience}, 11:114, 2017.

\bibitem{barak2017recurrent}
Omri Barak.
\newblock Recurrent neural networks as versatile tools of neuroscience
  research.
\newblock {\em Current opinion in neurobiology}, 46:1--6, 2017.

\bibitem{kriegeskorte2018cognitive}
Nikolaus Kriegeskorte and Pamela~K Douglas.
\newblock Cognitive computational neuroscience.
\newblock {\em Nature neuroscience}, 21(9):1148--1160, 2018.

\bibitem{lecun2015deep}
Yann LeCun, Yoshua Bengio, and Geoffrey Hinton.
\newblock Deep learning.
\newblock {\em nature}, 521(7553):436--444, 2015.

\bibitem{schmidhuber2015deep}
J{\"u}rgen Schmidhuber.
\newblock Deep learning in neural networks: An overview.
\newblock {\em Neural networks}, 61:85--117, 2015.

\bibitem{hassabis2017neuroscience}
Demis Hassabis, Dharshan Kumaran, Christopher Summerfield, and Matthew
  Botvinick.
\newblock Neuroscience-inspired artificial intelligence.
\newblock {\em Neuron}, 95(2):245--258, 2017.

\bibitem{cambria2014jumping}
Erik Cambria and Bebo White.
\newblock Jumping nlp curves: A review of natural language processing research.
\newblock {\em IEEE Computational intelligence magazine}, 9(2):48--57, 2014.

\bibitem{rahwan2019machine}
Iyad Rahwan, Manuel Cebrian, Nick Obradovich, Josh Bongard, Jean-Fran{\c{c}}ois
  Bonnefon, Cynthia Breazeal, Jacob~W Crandall, Nicholas~A Christakis, Iain~D
  Couzin, Matthew~O Jackson, et~al.
\newblock Machine behaviour.
\newblock {\em Nature}, 568(7753):477--486, 2019.

\bibitem{voosen2017ai}
Paul Voosen.
\newblock The ai detectives, 2017.

\bibitem{hutson2018artificial}
Matthew Hutson.
\newblock Artificial intelligence faces reproducibility crisis, 2018.

\end{thebibliography}
\end{document}